\documentclass[runningheads,citeauthoryear]{apinv}
\usepackage{epsfig,cite,graphics}
\usepackage{marvosym} %For astronomical symbols % % %
\usepackage[utf8]{inputenc}

\def\kms{km~s$^{-1}$}  
\newcommand{\msun}{$M_\odot$}

\begin{document}

\title{High-resolution optical spectroscopy of Nova~V392~Per}
\titlerunning{Spectroscopy of V392~Per}
\author{K. A. Stoyanov\inst{1}, T. Tomov\inst{2}, I. Stateva\inst{1}, S. Georgiev\inst{1}}
\authorrunning{K. Stoyanov et al.}
\tocauthor{K. Stoyanov et al.} 
% Command tocautor{} is used by the Latex to give author names 
% to the Contents of the volume (automatically generated)
\institute{Institute of Astronomy and National Astronomical Observatory, Bulgarian Academy of Sciences, 72 Tsarigradsko Chaussee Blvd., 1784, Sofia, Bulgaria  \and Centre for Astronomy, Faculty of Physics, Astronomy and Informatics, Nicolaus Copernicus University, Grudziadzka 5, 87-100 Torun, Poland
	\newline
	\email{kstoyanov@astro.bas.bg}    }
\papertype{Submitted on xx.xx.xxxx; Accepted on xx.xx.xxxx}	
% Papertype can be "Research report", "Review", "Invited lecture", "Conference talk", 
% "Conference poster", "Lecture at scientific seminar", "Summary of dissertation",  etc.
\maketitle

\begin{abstract}
Here we analyze high-resolution spectra of the Nova V392~Per obtained during the 2018 outburst. The H$\alpha$ and H$\beta$ emission lines show a triple-peak structure
with radial velocities of about -2000~\kms, -250~\kms\ and 1900~\kms\ respectively. The near infrared spectrum is dominated by the narrow and single-peaked Paschen lines of hydrogen and the O~I~$\lambda$8446 
and O~I~$\lambda$7773 emission lines. Using DIBs and the K~I line, we estimate the interstellar excess towards V392~Per. Based on AAVSO and ASAS-SN photometry data, we calculate that the 
t$_2$ and t$_3$ decline times are $\sim$ 3~d and $\sim$ 11~d respectively, which classifies 
V392~Per as a very fast nova. We also briefly discuss the similarity between V392~Per and other very fast novae and the possible future evolution of the system in terms of the hibernation model.\\
\end{abstract}
\keywords{novae, cataclysmic variables -- stars: individual: V392~Per}

\section*{Introduction}
In the cataclysmic variables, the nova outburst is powered by thermonuclear runaway on the surface of the white dwarf (Bode \& Evans 2008). Over long periods of time, the hydrogen-rich 
material is being accreted from the donor star and forms an envelope on the white dwarf. Once the critical temperature (a.k.a. Fermi temperature $\sim$ 7 $\times$ 10$^7$~K)
and the critical pressure of the accreted material are reached, the thermonuclear runaway is ignited, causing a dramatic brightening with maximum magnitude in the range of -5 to -10.7 (Shafter et al. 2009).
As a result of the thermonuclear runaway, the temperature in the nuclear burning region can reach $\sim$ 4 $\times$ 10$^8$~K for very massive white dwarfs (Starrfield, Iliadis \& Hix 2016).
It is confirmed by observations that the nova ejecta enriches the Interstellar Medium with heavy elements and dust (Gehrz et al. 1998).\\
Hundreds of close binaries in the Galaxy have been known to have undergo one nova outburst. They are called {\it classical novae}. The {\it recurrent novae} are objects that have experienced more 
than one nova
outburst (Mukai 2015). It is believed that all classical novae are recurrent with intervals between the outbursts of many centuries.
The {\it dwarf novae} are interacting binaries in which the late-type donor overfills its Roche lobe and transfers material to the white dwarf via an accretion disc. These objects display outbursts during 
which the system brightens by 2 to 5 magnitudes on timescales of days to weeks. The dwarf nova outbursts are driven by a thermal-viscous disc instability (Osaki 1974; Smak 1984).\\
The {\it hibernation model} connects the classical novae with the dwarf novae (Shara et al. 1986). According to the model, a system can change state through various types of CVs on its way into and out 
of hibernation as the accretion rate changes. After the nova outburst, the system should appear as a nova-like variable with a high accretion rate but no outbursts. As the accretion rate decreases,
the system should transit to a dwarf nova, and then finally go into hibernation state when the accretion stops. After a time period that could be of the order of thousands of years,
the gravitational radiation will return the system in a contact, the companion will overflow its Roche lobe causing a restart of the accretion. Then the system will eventually be back through the cycle 
from dwarf nova to a nova-like and back to another nova outburst (Pagnotta 2015). The {\it hibernation model} is supported by the discovery of nova shells around dwarf novae Z~Cam (Shara et al. 2007)
and AT~Cnc (Shara et al. 2012), and the long-term observations of V1213~Cen (Mr{\'o}z et al. 2016).\\
V392~Per is classified as a Z~Cam-type dwarf nova by Downes \& Shara (1993). On 2018 April 29.474 UT, Nakamura (2018) reported the discovery of a new transient that is coincident with the 
position of V392~Per. Wagner et al. (2018) confirm spectroscopically that this transient is a nova eruption in a Fe~II curtain phase. Therefore, V392~Per joins the group of a very few dwarf novae that subsequently
undergo a nova outburst. Later, Li, Chomiuk \& Strader (2018) detected a $\gamma$-ray emission from V392~Per.\\
In this paper we analyze high-resolution spectra of V392~Per, obtained in May 2018.

\section{Observations}
We observed V392~Per with the ESPERO fiber-fed echelle spectrograph (Bonev et al. 2017), mounted on the 2m RCC telescope at Rozhen National Astronomical Observatory (Bulgaria).
The two spectra were obtained on 2018 May 1 18.74 UT and May 2 18.80 UT. The spectra cover the range of 4400-9000~\AA\ with a resolving power of $\sim$30~000. The data were reduced in the standard 
way including bias removal, flat-field correction, wavelength calibration and correction for the Earth's motion. 
The processings and the measurements of the spectra are performed using standard routines provided by 
IRAF (Tody 1993).

\section{Results}
\subsection{Light curve}
\begin{figure}[!h]
	\begin{center}
		\centering{\epsfig{file=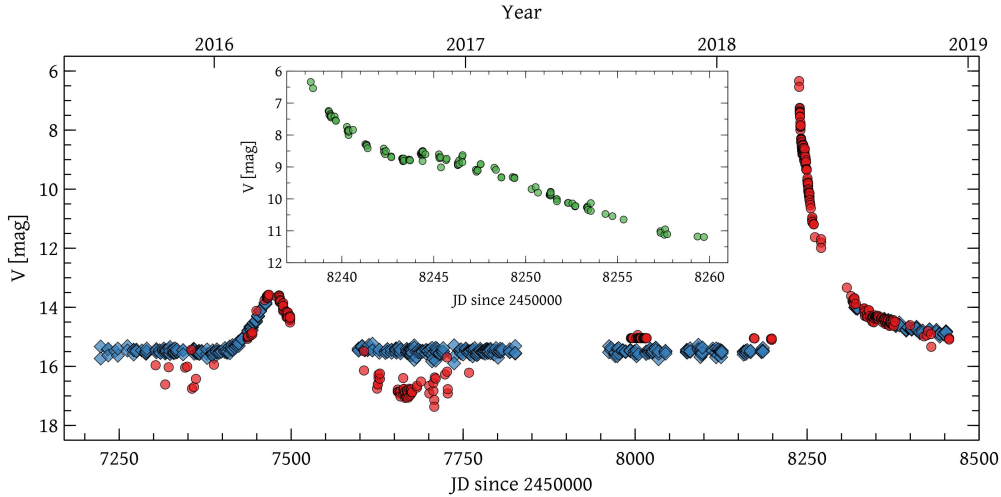, width=\textwidth}}
		\caption[]{The \mbox{V}-band light curve of V392~Per between July 2015 and December 2018, based on AAVSO (red dots) and ASAS-SN (blue diamonds) observations.
		The ASAS-SN data were corrected by 0.45 mag in order to adjust for an evident systematic off-set.
		The error bars are not shown because they are smaller than the used symbols in almost all cases.}
		\label{fig:lc}
	\end{center}
\end{figure}

The V-band light curve in the period July 2015 -- December 2018 based on AAVSO (Kafka 2018) and ASAS-SN data (Shappee et al. 2014; Kochanek et al. 2017) is shown on Fig.\ref{fig:lc}.
The light curve demonstrates the difference between the 2016 dwarf nova and the 2018 classical nova outbursts of V392~Per. The dwarf novae usually show outbursts of amplitudes of 4 -- 6 mag 
(Buat-M{\'e}nard, Hameury \& Lasota 2001). During the period 2004 -- 2018, the light curve of V392~Per shows a quiescent state with a V-band magnitude of 16 -- 17 mag, accompanied by three or 
four dwarf nova outbursts, the last one in 2016 (Darnley \& Starrfield 2018). The outburst on 2018 April 29 is the first observed nova eruption from this system. According to the AAVSO light curve,
the maximum brightness of V392~Per is V=6.34~mag. These observations suggest an eruption amplitude of $\sim$10 -- 11 magnitudes, which could indicate the presence of an evolved donor in the system.  
Using the interstellar excess of E(B-V)=1.18$\pm$0.10~mag (see Sect.~\ref{spectra}) and the distances to V392~Per 3416$^{+750}_{-533}$~pc (Bailer-Jones et al., 2018) and 4161$^{+750}_{-533}$~pc (Shaefer 2018),
we can derive the absolute V-band magnitude at maximum as -10~mag and -10.4~mag respectively.
The two values are relatively close to each other.\\
The times of decline by 2 and 3 magnitudes from the peak (t$_2$ and t$_3$) for V392~Per are $\sim$ 3~d and $\sim$ 11~d, respectively, which classifies the system as a very fast nova (Payne-Gaposchkin 1957).
The observations of the very fast novae V2672~Oph and U~Sco show that the t$_3$ decline time is two times longer than the t$_2$ time (Munari et al., 2011). The plateau in their light curves begins at
4 -- 6 magnitudes below the maximum. In the case of V392~Per, t$_3$ time 
is almost four times longer than the t$_2$ time. As it can be seen on Fig.~\ref{fig:lc}, the t$_3$ time coincides with the plateau in the light curve of V392~Per. In this case, the plateau begins at
about 2.5~mag below the maximum and this extends the t$_3$ time to about 11 days. The possible explanation of this fact is that the maximum of the outburst has been probably  missed.\\
Using the relations between the M$_V$ and the t$_2$ and t$_3$ decline times from Downes \& Duerbeck (2000), we estimate the M$_V$ at maximum to be $\sim$ -10.1~mag and -9.3~mag, respectively. 
Selvelli \& Gilmozzi (2019) published new maximum magnitude versus rate of decline (MMRD) relations between M$_V$ and t$_3$ time on the base of 18 novae with very well known characteristics and the GAIA
DR2 distances. Using their relations, we obtain an absolute magnitude at maximum $\sim$ -8.9~mag. The difference between these three values can be another proof that supports the idea that the nova outburst is detected after the brightness maximum.

\subsection{Spectra}
\label{spectra}

\begin{figure}[!h]
	\begin{center}
		\centering{\epsfig{file=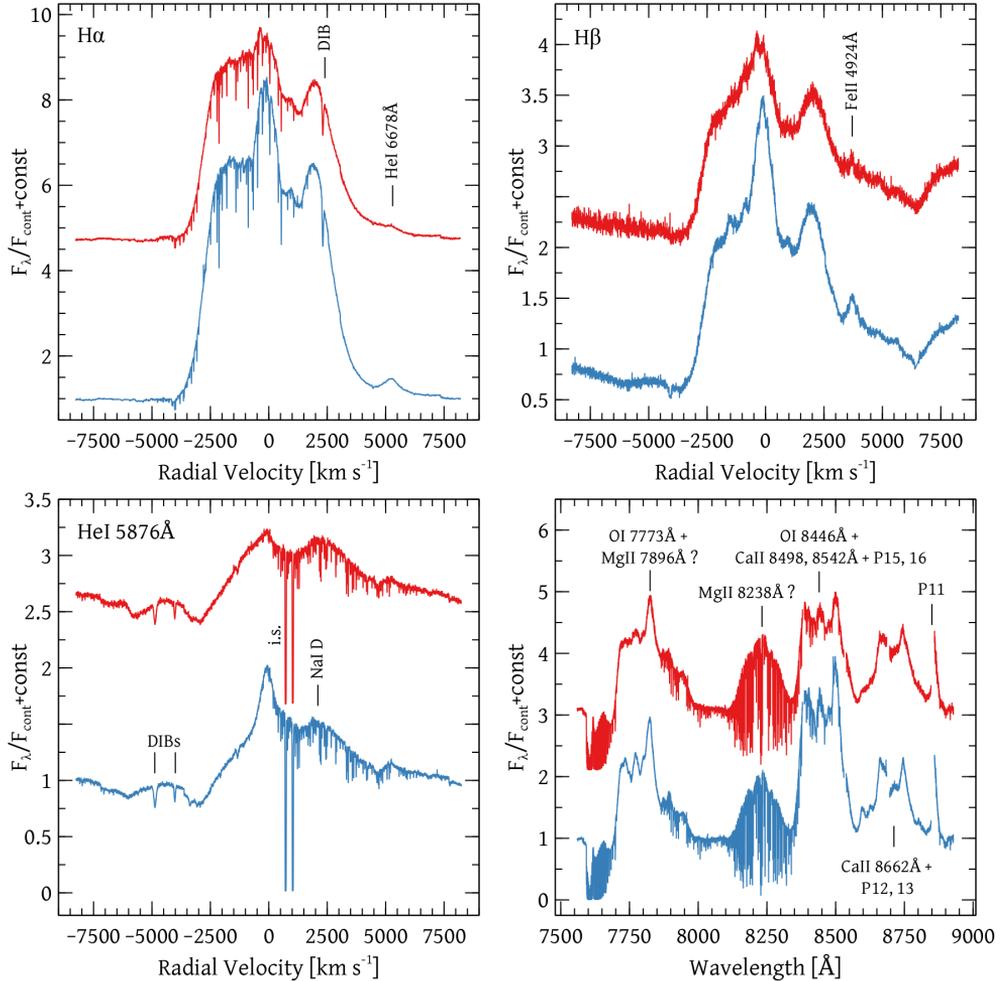, width=\textwidth}}
		\caption[]{Comparison of selected profiles in the V392~Per spectra obtained shortly after the maximum brightness on 2018, May 1 (upper profiles in all panels) and May 2 
		(bottom profiles in all panels). The sharp absorption features visible in the spectra are telluric lines, with the exception of the NaI D interstellar lines and the DIBs 
		at 5780\,\AA, 5797\,\AA\ and 6614\,\AA. In the red part of the spectrum in the lower right panel the inter-order gaps are also visible.}
		\label{fig:sp}
	\end{center}
\end{figure}
The Diffuse Interstellar Bands (DIBs) at 5780\AA, 5797\AA\ and 6614\AA, and the interstellar lines Na~I and K~I are very prominent in our spectra. The interstellar lines show two components with 
radial velocities of -23.7$\pm$1.5~\kms\ and 3.5$\pm$0.3~\kms\ respectively. Using the relations of Munari \& Zwitter (1997) for K~I and of Puspitarini, Lallement \& Chen (2013) for the DIBs, 
we derived an interstellar excess E(B-V)$\sim$1.18$\pm$0.10~mag. The Na~I lines are saturated and we cannot use them for any measurements.\\
Fig.~\ref{fig:sp} shows selected lines in the spectra of V392~Per -- the H$\alpha$ and H$\beta$ Balmer lines, the HeI~5876\AA\ emission line and the region around O~I~$\lambda$8446 and 
O~I~$\lambda$7773 emission lines. The spectra are dominated by very broad emissions with a complex shape. 
The strongest features in our spectra are H$\alpha$, H$\beta$, FeII 42 multiplet, CaII~IR triplet, and O~I~$\lambda$7773 and $\lambda$8446. These very broad lines are heavily blended. 
He~I lines 5876\AA\ and 6677\AA\ are relatively weak with an increase of the emission intensity during the second night. The FWHM measured for the H$\alpha$ and H$\beta$ emission lines is 
5600$\pm$200~\kms. They have three-component profiles with radial velocities of about -2000~\kms, -250~\kms\ and 1900~\kms\ respectively. Note that the central emission in the H$\alpha$ and
H$\beta$ lines also increases significantly during the second night. Weak P~Cyg absorption components are also visible in 
their profiles. We measured the radial velocity to be about -4000~\kms\ for one absorption component of H$\alpha$ and about -3700~\kms\ and -4100~\kms\ for two absorptions of H$\beta$. 
Three very weak absorptions of He~I 5876\AA\ with radial velocities of about -1350~\kms, -3035~\kms\ and -3350~\kms\ are also visible in the spectrum.\\
In July 2018 V392~Per enters a post-nova eruption nebular phase (Darnley 2018). The H$\alpha$ and H$\beta$ lines remain with a similar three-component structure but the intensity 
of the [O~III]~$\lambda$4363, $\lambda$4959 and $\lambda$5007, and He~II~$\lambda$4686 increases extremely. More dramatic changes in the spectrum of V392~Per occured in August 2018, when the nova turns 
into a neon nova (Munari \& Ochner 2018). In the spectra obtained during that time, all emission lines are composed by two broad and well separated components with a stronger blue component and a third central narrow
emission component. Moreover,
the [Ne~V]~$\lambda$3426 line appears comparable in intensity to the H$\alpha$ line.\\
The infrared part of the spectra of V392~Per is dominated by the Paschen lines of the hydrogen and the O~I~$\lambda$8446 and O~I~$\lambda$7773 emission lines (see the lower right panel on Fig.~\ref{fig:sp}).  
The profiles of the Paschen lines appear to be narrower and with single-peak structure in comparing to the Balmer lines. It seems that several emission lines of ionized calcium and 
magnesium can be identified in the spectra -- probably Mg~II~$\lambda$8238 and Mg~II~$\lambda$7896, blended with an O~I~$\lambda$7773 emission line [see Table 2 in Williams (2012)]. In any case,
the identification of the emission lines in that part of the spectrum is difficult because of the very large expansion velocities and the irregular profiles of the lines that cause large blending.\\

\section{Discussion and concluding remarks}
V392~Per is classified as a He/N nova, which is usually a distinctive for fast novae (Williams 2012). The He/N novae have V$_{ej}$ $>$ 2500~\kms\ (Bode \& Evans 2008). 30\% of all novae with V$_{ej}$ $>$ 
2500~\kms\ evolve into a neon nova, which is the case of V392~Per (Munari \& Ochner 2018). The object has very similar observational properties to the recurrent novae V2672~Oph and U~Sco 
(Munari et al. 2011). The donor stars in U~Sco and V2672~Oph are main sequence stars (Hanes 1985; Munari et al. 2011), in contrary to the symbiotic recurrent novae RS~Oph and T~CrB which contain
red giants as donors (Anupama \& Miko{\l}ajewska 1999; M{\"u}rset \& Schmid, 1999). Darnley \& Starrfield (2018) performed SED analysis of V392~Per and compared it to other recurrent novae.
The results support the sub-giant nature of the donor star and confirm the similarity between V392~Per and the recurrent novae U~Sco and M31N~2008-12a.\\ 
V392~Per and U~Sco are long-period systems -- 
P$_{orb}$ of U~Sco is 1.23~d (Schaefer \& Ringwald 1995) and the light curve during the outburst of V392~Per also suggests a long orbital period. Mukai (2015) proposed that the long-period novae
are undergoing thermal-timescale mass-transfer or did it at some point of their evolution (Schenker et al. 2002), which could explain the recurrent nova outbursts in these systems.\\
If V392~Per is a recurrent nova, the white dwarf should be close to the Chandrasekhar limit and it should accrete hydrogen-rich material at very high rates in order to build a critical-mass 
envelope quickly enough to secure the recurrent nova outbursts (Yaron et al. 2005). Selvelli \& Gilmozzi (2019) give correlations between the t$_2$ and t$_3$ decline times and different parameters 
of the white dwarf. Using the correlation between t$_2$ and t$_3$ and the mass of the white dwarf, we obtain masses $\sim$ 1.21~\msun\ and $\sim$ 1.08~\msun\ with average value 
M$_{WD}~\sim$ 1.15$\pm$0.07~\msun. Using the t$_3$ decline time, we can calculate the mass accretion rate -- 
$\dot{M}_{WD}$ $\sim$ 1.02 $\times$ 10$^{-9}$~\msun~yr$^{-1}$.\\ 
As we pointed earlier, the spectral behavior of V392~Per is very similar to other very fast novae such as U~Sco and V2672~Oph. Munari \& Ochner (2018) find that emission line profiles 
of V392~Per during the neon nova stage are similar to those observed at late stages in the evolution of the very fast and He/N nova V2672~Oph. Munari et al. (2011) proposed that they are formed
as a bipolar flows or blobs aligned with the line of sight and an equatorial ring-like structure seen face-on. The observational evidence points to a similar model for V392~Per as well.\\
As for the future evolution of V392~Per, there are two possible scenarios. According to the {\it hibernation model}, the system can appear as a nova-like variable star with a high accretion rate but no 
outbursts. As the accretion rate decreases, the system should transform into a dwarf nova. A similar evolution model is proposed for the dwarf nova BK~Lyn that transformed from nova through a nova-like and 
finally into a dwarf nova (Patterson et al. 2013). Another possibility is displaying recurrent nova outbursts with a recurrence time $<$100~yr, before entering hibernation.\\
In any case, optical and spectral follow-up observations of V392~Per in the next decades will be much valuable for testing the cataclysmic variables and nova evolution.

{\bf Acknowledgments:} 
We thank the referee, Prof. Radoslav Zamanov, for his comments and suggestions which improved the final quality of this article.
This work is supported by grants K$\Pi$-06-H28/2018, DN~08-1/2016, DN~18-10/2017 and DN~18-13/2017 of the Bulgarian National Science Fund and it is a part of
the joint project ``Spectral and photometric study of variable stars'' of the Bulgarian Academy of Sciences and the Polish Academy of Sciences.\\
The authors acknowledge the variable star observations from the AAVSO International Database contributed by observers worldwide and used in this research. 
ASAS-SN is supported by the Gordon and Betty Moore
Foundation through grant GBMF5490 to the Ohio State
University and NSF grant AST-1515927. Development of
ASAS-SN has been supported by NSF grant AST-0908816,
the Mt. Cuba Astronomical Foundation, the Center for Cosmology and AstroParticle Physics at the Ohio State University, the Chinese Academy of Sciences South America Center
for Astronomy (CAS- SACA), the Villum Foundation, and
George Skestos.

\end{document}